\documentclass[aps,prl,twocolumn,groupedaddress,superscriptaddress,showpacs]{revtex4-1}
\usepackage{graphicx}
\usepackage{dcolumn}

\begin{document}

\title{Mimicking Nanoribbon Behavior Using a Graphene Layer on SiC}

\author{Matheus P. Lima}
\email[]{mplima@if.usp.br}
\affiliation{Instituto de F\'isica,
Universidade de S\~ao Paulo, CP 66318, 05315-970, S\~ao Paulo, SP,
Brazil}

\author{A. R. Rocha}
\affiliation{Centro de Ci\^encias Naturais e Humanas,
Universidade Federal do ABC, Santo Andr\'e, SP, Brazil}

\author{Ant\^onio J. R. da Silva}
\email[]{ajrsilva@if.usp.br}
\affiliation{Instituto de F\'isica,
Universidade de S\~ao Paulo, CP 66318, 05315-970, S\~ao Paulo, SP,
Brazil}
\affiliation{Laborat\'orio Nacional de Luz S\'{\i}ncrotron, CP 6192, 13083-970, Campinas, SP, Brazil}

\author{A. Fazzio}
\email[]{fazzio@if.usp.br}
\affiliation{Instituto de F\'isica,
Universidade de S\~ao Paulo, CP 66318, 05315-970, S\~ao Paulo, SP,
Brazil}

\date{\today}

\begin{abstract}
We propose a natural way to create quantum-confined regions in graphene in a system that allows large-scale device integration. We show, using first-principles calculations, that a single graphene layer on a trenched region of $[000\bar{1}]$ $SiC$ mimics i)the energy bands around the Fermi level and ii) the magnetic properties of free-standing graphene nanoribbons. Depending on the trench direction, either zigzag or armchair nanoribbons are mimicked. This behavior occurs because a single graphene layer over a $SiC$ surface loses the graphene-like properties, which are restored solely over the trenches, providing in this way a confined strip region.
\end{abstract}


\maketitle

Graphene has attracted enormous attention due to its potential for application in devices at the nanometer scale. The very high mobility of graphene charge carriers, exceeding $200.000~cm^2/V.s$, presents a possibility of fabrication of field effect transistors (FETs) as an alternative to $Si$-based devices \cite{rise}. As is
well known, graphene has a linear dispersion relation with valence and conduction band degenerate at the so called Dirac point \cite{castroRMP}. This electronic behavior implies that, although pristine graphene can be used for applications in radio frequency (RF) devices \cite{100GHz}, it has no energy gap. Consequently,
it presents a very small current on/off ratio for transistors \cite{sgt}. Thus, one of the most fundamental challenges for researchers in the field is to fabricate graphene-like structures with a band gap that i) preserve the above mentioned qualities; ii) the band gap is fabricated in a controllable way and iii) in such a way that it is possible to integrate them in large-scale circuits.

A possible solution for these challenges would be to use graphene nanoribbons (GNR) \cite{kim,engeanering}.
Huge efforts have been made with the aim of understanding the properties of GNRs, and possible applications on electronics, spintronics, memory devices and sensors have been proposed in the literature \cite{gnr-prop1,gnr-prop2,gnr-prop3,gnr-prop4,gnr-prop5}. The gap can be controlled by the width of the ribbon and the high mobility is still preserved in these systems. Considering large-scale circuits and the perspective of use of graphene on wafers for industrial applications, the process of obtaining a single layer of graphene via heating a hexagonal SiC surface until the sublimation of Si surface atoms (with temperatures over $1000\,^{\circ}C $)\cite{berger} is attracting attention as a viable option. Recently, there have been proposals for FETs which are based on graphene layers grown on SiC substrates \cite{bitrans,TSiC1,TSiC2}.

\begin{figure}[h!]
\includegraphics[width=8.2cm]{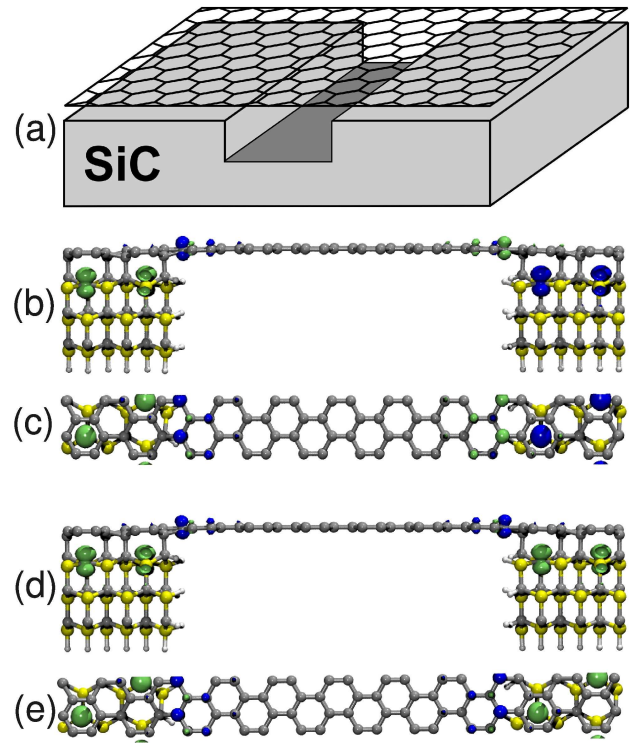}%
\caption{\label{fig1} (Color online) (a) Schematic picture of  a single graphene layer over a SiC surface containing a trench. (b-e) Side and top views of the local magnetization ($\rho_\uparrow(\vec{r})-\rho_\downarrow(\vec{r})$)
for a graphene sheet over SiC with a $20.6\AA$ wide trench cut along the $\Gamma - K$ direction. (b-c) Anti-ferromagnetic, and (d-e) ferromagnetic configuration. Dark (blue) indicates excess of $\uparrow$ electrons, while the light (green) indicate excess of $\downarrow$ electrons.}
\end{figure}

\begin{figure*}
\includegraphics[width=17cm]{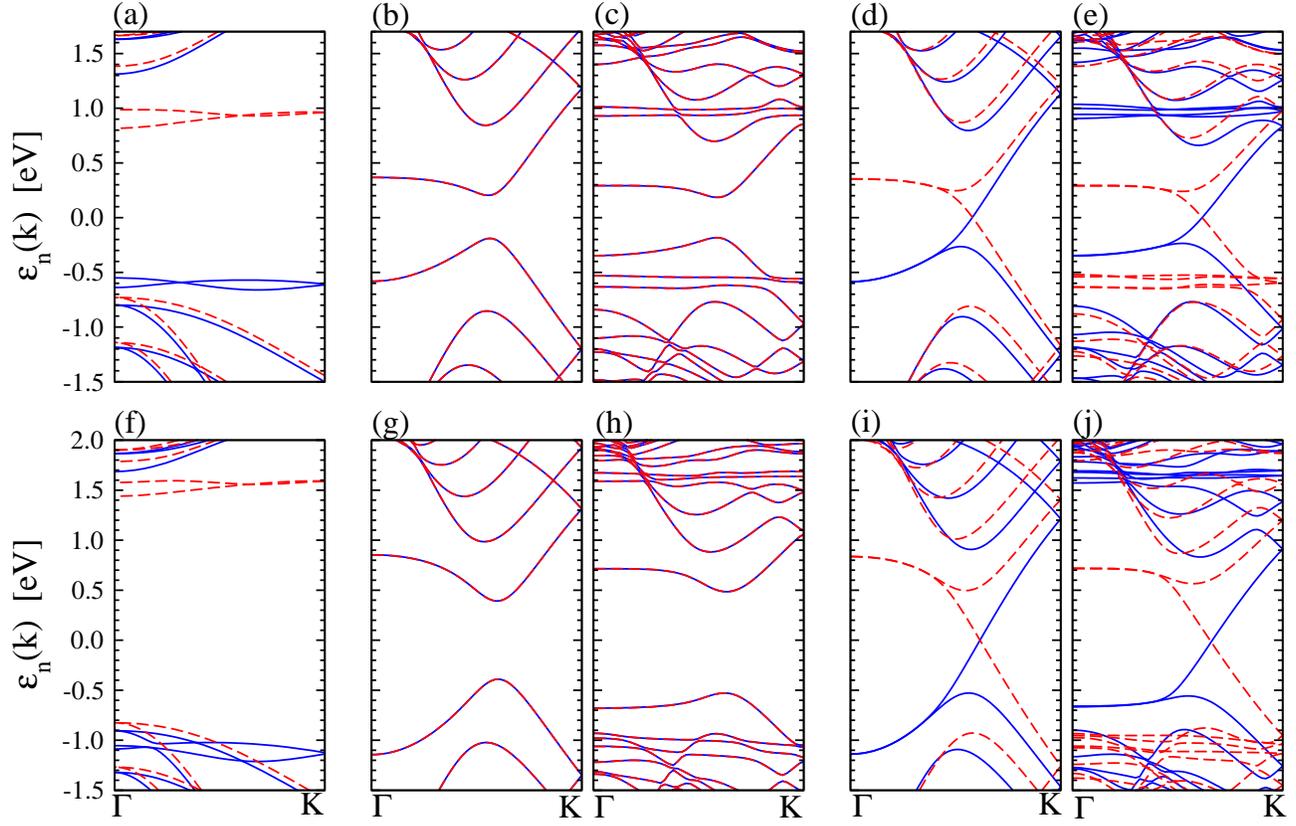}
\caption{\label{fig2} (Color Online) Energy bands.  
(a) Single sheet of graphene over a crystalline surface of $SiC$ without a trench.  
Anti-ferromagnetic configuration: 
(b) FS-zGNR and (c) single sheet of graphene over $SiC$ with a trench in the $\Gamma - K$ direction.
Ferromagnetic configuration:
(d) FS-zGNR and (e) graphene over trenched $SiC$. 
(f-j) Self-interaction correction for cases reported in (a-e). The trenches considered here have a 
width of $20.6\AA$. The solid (blue) lines for the $\uparrow$ states, and the dashed (red) lines for the $\downarrow$ states.}
\end{figure*}

In this letter we propose a system that naturally creates graphene nanoribbons on top of SiC substrates.
We show, using {\it ab-initio} density functional theory (DFT)\cite{dft,methodology}, that a single sheet of graphene over the $[000\bar{1}]$ (carbon rich) surface of SiC can mimic a free-standing graphene nanoribbon (FS-GNR) in places where substrate atoms are removed, creating in this way trenches in the SiC \cite{trench1,trench2, trench3}. In Fig. \ref{fig1}(a) a schematic picture of our proposal is depicted. Depending on the direction of the trench, either FS-GNRs with zig-zag (FS-zGNRs) or armchair edge shapes (FS-aGNRs) are mimicked.  When the trench is in the $\Gamma-K$ graphene direction, ferromagnetic (metallic) and anti-ferromagnetic (semi-conductor) solutions are possible, exactly as one would find in the FS-zGNRs. For the trench along the $\Gamma-M$ graphene direction, tree families of energy bands can be identified and associated to the $3p$, $3p+1$ and $3p+2$ families of FS-aGNRs, where $p$ is a positive integer \cite{louie}. For graphene over the $[0001]$ (silicon rich) SiC surface, the half-filled dangling-bond states pin the Fermi energy at $\approx 0.5eV$ above the Dirac point \cite{pinning} (near the bottom of the conduction band), preventing the appearance of states that mimic FS-GNR near the Fermi energy, even in the presence of a trench.

In our geometry,  a single layer of graphene is positioned over SiC surfaces with the $\sqrt{3}\times \sqrt{3}R30\,^{\circ}$ reconstruction. The substrate we consider has 3 bilayers of SiC in the ABC stacking pattern, representing a surface-terminated thin slab of 4H-SiC with periodic boundary conditions. In this situation, there are covalent bonds between the graphene and the top-surface atoms that destroy the free-standing graphene
properties (such as the conic dispersion around the Fermi-level), in agreement with previous {\it ab-initio} simulations\cite{mattausch,varchon} as well as experiments performed in both C-rich \cite{hass} and Si-rich surfaces \cite{Si-rich}. We obtain the trench by removing atoms of the substrate from underneath the graphene layer in the central region, considering two possible directions, namely along the $\Gamma-K$ and $\Gamma-M$ graphene Brillouin zone directions. Along the $\Gamma-K$ direction, trenches $7.5$, $11.9$, $20.6$ and $34.8~\AA$ wide were considered, whereas for the $\Gamma-M$ direction, we consider trenches $5.2$, $7.7$, $10.4$, $13.0$, $15.7$, $18.2$, $20.9$, $23.5$ and $26.2~\AA$ wide.

In Fig. \ref{fig2} we show the energy bands along the $\Gamma-K$ direction. (The Fermi energy is shifted to $0.0~eV$ in all graphs). Fig. \ref{fig2}(a) represents the dispersion relation for the system without a trench.
The two flat bands above and below the Fermi-level are associated with dangling bonds (DB) localized on the carbon atoms on the SiC surface. We notice a splitting between the majority ($\uparrow$) and minority spin
($\downarrow$) DB states. In Fig. \ref{fig2} (b) we show the band structure for a FS-12-zGNR,
where the convention of Ref. \onlinecite{louie} is adopted. In Fig. \ref{fig2}(c) we present the dispersion relation for a system with a trench $20.6\AA$ wide. In this situation, the free-standing carbon atoms form a region equivalent to a FS-12-zGNR. The edge magnetism present in the FS-zGNR \cite{martins-tb} is also present in the system with the trench. One can find both an anti-ferromagnetic (AF) configuration, and a ferromagnetic (FM) one. Similarly to the free standing nanoribbons, we also find that the AF has a lower energy than the FM configuration, and the energy differences decay quickly with the width of the ribbon, again as obtained in FS-zGNR. For the AF case, we compare the energy bands of a FS-12-zGNR with the dispersion relation for the system with the trench (shown in the Fig. \ref{fig2}(b-c)). The main characteristics, {\it i.e.}, the presence of a gap of similar value, and the shape of the bands around the Fermi level, are very close in both systems. The DB levels that are present in the system without the trench still appear even in the presence of a trench, since their origin comes from the non-free-standing region. In Fig. \ref{fig2}(d-e), the corresponding bands for the same systems, but when a ferromagnetic configuration is adopted, are shown. One can note that as it occurs for the FS-zGNR, the
system with a trench becomes metallic.

The geometry and the local magnetization for the AF case are presented in Fig. \ref{fig1}(b-c) - side and the top views, respectively - where one can identify DB states with a non-zero local magnetization as well as local magnetic moments on the the free-standing graphene atoms located just above the wall of the trench, which mimic the edge atoms when compared to FS-zGNRs. An analogous behavior can be noted in the FM configuration (Fig. \ref{fig1}(d-e)), however with a reversal of the magnetization on one of the edges. Thus, similarly to what is observed in FS-zGNRs, the magnetic ordering (AF or FM) is defined by the local magnetic moments that appear at the graphene edges, above the wall of the trench, precisely the sites that mimic the edge sites of FS-zGNRs. Moreover, the local magnetization coming from the DB states, that appear between the substrate and the graphene, can be flipped up or down without changing the semi-conductor or the metallic character of the AF and FM states, respectively. However, we find a small magnetic coupling between the free-standing region edge sites and their nearest DB, that favors an anti-ferromagnetic configuration (if one DB is flipped an increase of $0.04eV$ in the total energy occurs).

It is well know that in systems with localized states, which is the case investigated in this work\cite{tightbinding}, the self-interaction correction \cite{zunger} is important to obtain a better quantitative description of the energy bands. Thus, to include such correction, we implement the methodology proposed by Filippetti {\it et al.} \cite{filippetti} in the SIESTA code\cite{self-int}. In Fig. \ref{fig2} (f) we show the energy bands including the self-interaction correction for the graphene on a SiC substrate without a trench. Comparing with the same system without this correction, (shown in Fig. \ref{fig2}(a)), one can see that the splitting between the $\uparrow$ and $\downarrow$ DB states increases from $1.5~eV$ to $2.6~eV$. Without the self-interaction correction, the highest occupied state is a DB state, whereas after the inclusion of such correction it becomes a substrate bulk state. Other effects of this correction include the increase of the local magnetization at the edge atoms of FS-zGNRs (from $0.28$ to $0.37~\mu_B$), and a considerable increase in the gap. In Fig. \ref{fig2}(h) we present the energy bands for the system with a trench in the $\Gamma-K$ direction with the inclusion of the self-interaction correction for the AF configuration. This calculation can be well compared with an AF FS-12-zGNR, shown in Fig. \ref{fig2} (g), where the self-interaction correction is also employed.

\begin{figure}[!t]
\includegraphics[width=8.5cm]{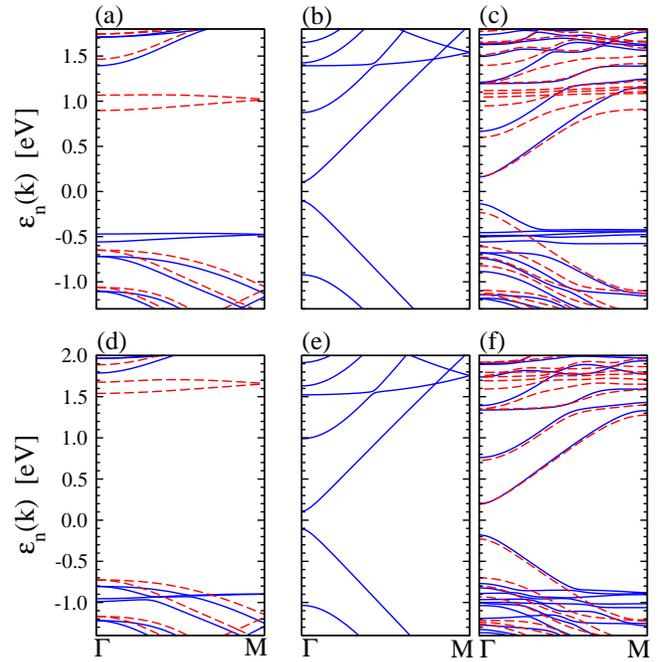}
\caption{\label{fig3} (Color Online) Energy bands in the $\Gamma-M$ graphene direction.
(a) Single sheet of graphene over a crystalline SiC surface. 
(b) Armchair  FS-GNR with 7 hexagons from one edge to another.
(c) Graphene layer over SiC containing
a trench $18.2~\AA$ wide along the $\Gamma-M$ graphene direction. 
(d-f) Self-Interaction-Correction calculaions for the same cases presented
in (a-c). The solid (blue) lines correspond to the $\uparrow$ states, and the dashed (red) lines are
for the $\downarrow$ states . }
\end{figure}
If we now consider the trench along the $\Gamma-M$ graphene direction, the system mimics an armchair FS-GNR, as can be seen in the Fig. \ref{fig3}. For the sake of comparison we plot the dispersion relation for a single layer of graphene over a $SiC$ surface without the presence of a trench along the $\Gamma-M$ direction. In Fig. \ref{fig3}(b) we show the band structure for a FS-15-aGNR (following the notation of Ref. \onlinecite{louie}). In \ref{fig3}(c) we show the energy bands when the graphene is over a $SiC$ surface containing a trench of width $18.2~\AA$.
The flat bands in Figs. \ref{fig3}(a) and (c) at energies $-0.5~eV$ and $1.0~eV$ are due to the DB on the surface of the $SiC$. The energy bands presented in Figs. \ref{fig3}(b) and (c) are very similar to each other around the Fermi energy. The gap for the FS-aGNR is slightly smaller, and the graphene layer on $SiC$ presents a small spin splitting of the valence band. The energy bands shown in Fig. \ref{fig3}(d-f) include the self-interaction correction, and uses the same geometries of Figs. \ref{fig3}(a), (b) and (c), respectively. Interestingly, it has no effect on the gap; the DB states are essentially moved away from the Fermi level in such a way as to restore the spin degenerate system and to bring the behavior of the trenched graphene nanoribbon closer to the free-standing case.

\begin{figure}
\includegraphics[width=8.5cm]{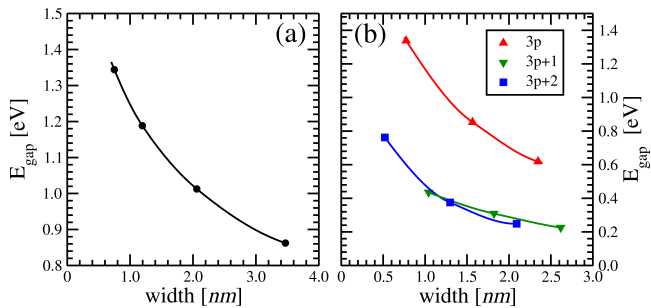}%
\caption{\label{fig4} Energy gaps. In (a) the zig-zag case, and in (b) the armchair case.
In both graphs calculations using the Self-Interaction-Correction are considered.}
\end{figure}

We have also calculated the gap dependence with the width of the trenches. In \ref{fig4}(a) we present the results along the $\Gamma-K$ graphene direction. The free-standing regions of these systems are similar to the zigzag FS-$N$-GNRs, with $N=\left\{4,6,12,16\right\}$. In this case, the energy gap monotonically decreases with the width of the trench, in a similar fashion to what occurs in the free-standing systems \cite{louie}. In Fig. \ref{fig4}(b) we consider the $\Gamma-M$ graphene direction. Similarly to what happens in the FS-aGNRs, 3 families of systems, named $3p$, $3p+1$ and $3p+2$, can be identified.
Since the graphene layer is strongly attached to the substrate, once the trench direction is defined in 
the SiC surface, there is no possibility for any further rotation of the graphene layer.
Thus, the orientation of the trench relativelly to graphene uniquely defines the properties of the ribbons, 
and intermediate orientations of the trenches will lead to edges that have a mixture of zigzag and armchair ribbons.

Summarizing, a single graphene layer on a trenched $[000\bar{1}]$ $SiC$ surface mimics the energy bands around the
Fermi level as well as the magnetic properties of FS-GNRs. With the trench along the $\Gamma-K$ graphene direction, we find a semiconductor (metallic) state with an AF (FM) order, exactly as it occurs in the FS-zGNRs. With the trench along the $\Gamma-M$ graphene direction, we find 3 families of systems, similarly to FS-aGNRs. This behavior occurs because a single graphene layer over a $SiC$ surface loses the graphene-like properties, and the presence of the trench restores the graphene-like properties only in a confined strip region. In this way, we propose a natural way to create quantum-confined regions in graphene in a system that allows large-scale device integration.

The authors acknowledge the financial support from the Brazilian funding agencies FAPESP and CNPq.


\begin{thebibliography}{99}
\bibitem{methodology} We use the SIESTA code\cite{siesta} with the Troullier-Martins pseudopotential
parametrization\cite{t-m}, a mesh cuttoff of 170 Ry for the real space grid integration,
and a k-sampling grid of $1\times1\times7$, where the 7 k-points are taken along the trench
direction. The GGA exchange-correlation functional is employed with the PBE parametrization\cite{pbe}.
All geommetries were fully relaxed within a force criterion of $0.03 eV/\AA$, and a vacum interval of $15~\AA$
between the slabs is used to avoid undesirable interactions between the periodic images.
\end{thebibliography}

\end{document}